\documentclass[conference,10pt]{IEEEtran}
\IEEEoverridecommandlockouts
\usepackage[utf8]{inputenc}

\usepackage{tikz,pgfplots}
\usetikzlibrary{positioning, calc,backgrounds}
\usepgfplotslibrary{fillbetween} 
\usetikzlibrary{patterns} 
\pgfplotsset{compat=1.18}
\usepackage{pgfplotstable}
\usepackage{ifthen} 
\usepackage{xcolor}
\definecolor{mygreen}{HTML}{A1A9D0} 
\definecolor{myyellow}{HTML}{F0988C} 
\definecolor{myorange}{HTML}{9E9E9E} 
\definecolor{myred}{HTML}{F93827}
\usepackage{tcolorbox}
\usetikzlibrary{arrows.meta}
\tcbuselibrary{most}
\usetikzlibrary{shadows}
\usetikzlibrary{shapes}
\usetikzlibrary{shapes.arrows}
\usetikzlibrary{shapes.geometric}
\usepgfplotslibrary{groupplots}

\usepackage{booktabs}
\usepackage{graphicx} 
\usepackage{bm}
\usepackage{amsmath}
\usepackage{amsfonts}
\usepackage{algorithm2e}
\usepackage{algpseudocode}
\usepackage{multirow}
\usepackage{amssymb}
\usepackage{pifont}
\usepackage{subfigure}
\usepackage{soul}
\usepackage{tabularx}
\newcolumntype{Y}{>{\centering\arraybackslash}X}
\newcolumntype{K}[1]{>{\centering\arraybackslash}m{#1}}
\newcolumntype{L}[1]{>{\centering\arraybackslash}p{#1}}
\usepackage{multirow}
\usepackage{xcolor}
\usepackage{marginnote}
\usepackage{subfigure}
\usepackage{cite}
\usepackage{bm}
\usepackage{mathtools}
\usepackage{hyperref}
\usepackage{ragged2e} 

\IEEEaftertitletext{\vspace{-1.5em}}


\newcommand{\RabiOmega}{\Omega} 
\newcommand{\dipole}{\bm{\mu}_{\text{eg}}} 
\newcommand{\Efield}{\bm{\mathcal{E}}} 

\author{
Zihang Song, Qihao Peng, Pei Xiao, Bipin Rajendran and Osvaldo Simeone
\thanks{Zihang Song, Bipin Rajendran and Osvaldo Simeone are with the Centre for Intelligent Information Processing Systems (CIIPS), Department of Engineering, King’s College London (KCL), London WC2R 2LS, U.K. (e-mail: \{zihang.song, bipin.rajendran, osvaldo.simeone\}@kcl.ac.uk). The work at KCL is supported in part by the European Union's Horizon Europe project CENTRIC (101096379), the EPSRC project (EP/X011852/1), and the Open Fellowships of the EPSRC (EP/W024101/1 and EP/X011356/1).}
\thanks{Qihao Peng and Pei Xiao are with Institute for Communication Systems (ICS), Home for 5GIC \& 6GIC, University of Surrey, Guildford, Surrey, GU2 7XH, U.K. (Email: \{q.peng, p.xiao\}@surrey.ac.uk).}
}

\DeclareMathAlphabet\mathbfcal{OMS}{cmsy}{b}{n}
\usepackage{titlesec}

\begin{document}
\title{CSI-Free Symbol Detection for Atomic MIMO Receivers via In-Context Learning}
\maketitle

\begin{abstract}
Atomic receivers based on Rydberg vapor cells as sensors of electromagnetic fields offer a promising alternative to conventional radio frequency front-ends. In multi-antenna configurations, the magnitude-only, phase-insensitive measurements produced by atomic receivers pose challenges for traditional detection methods. Existing solutions rely on two-step iterative optimization processes, which suffer from cascaded channel estimation errors and high computational complexity. We propose a channel state information (CSI)-free symbol detection method based on in-context learning (ICL), which directly maps pilot-response pairs to data symbol predictions without explicit channel estimation. Simulation results show that ICL achieves competitive accuracy with {higher computational efficiency} compared to existing solutions.
\end{abstract}


\vspace{-2mm}
\section{Introduction}
\vspace{-1mm}
Atomic receivers based on Rydberg vapor cells have emerged as a promising alternative to conventional radio frequency (RF) front-ends \cite{gong2024rydberg}. Traditional RF architectures rely on amplifiers, mixers, and filters, which introduce noise and limit sensitivity, especially at high carrier frequencies. In contrast, atomic receivers exploit quantum effects to directly sense electromagnetic (EM) fields, achieving high sensitivity and wide bandwidth without the need for frequency downconversion~\cite{sedlacek2012microwave,bussey2022quantum}. These advantages position atomic receivers as strong candidates for the acquisition of weak or noise-sensitive signals in next-generation wireless systems.

Unlike conventional architectures, atomic receivers do not produce complex baseband samples. Instead, signal acquisition is performed via optical readout mechanisms, which measure changes in the atomic population induced by the incident field. This process yields only the \emph{magnitude} of the total EM field intensity. As a result, the received signal follows a nonlinear magnitude-only non-coherent model that discards phase information and mixes contributions from all transmitters. This fundamental deviation from the linear model assumed in standard multiple-input multiple-output (MIMO) systems renders both channel-estimation-based coherent detection and channel-agnostic noncoherent methods, which typically assume linearity~\cite{ngo2025noncoherent}, unsuitable for atomic receivers.

To handle the nonlinear and phase-insensitive measurement channel model induced by atomic receivers, existing approaches \cite{cui2025towards,xu2025channel} often adopt a methodology consisting of channel state information (CSI) acquisition from pilot measurements followed by symbol detection based on the estimated channel. This framework faces two main limitations. First, the detection performance is inherently constrained by the performance of channel estimation; any mismatch or error in the first stage propagates to the second, degrading detection accuracy \cite{xu2025channel}. Second, the iterative optimization procedures involved in both steps introduce considerable computational overhead and latency. In some cases, particularly for algorithms that decouple operations across antennas \cite{cui2025towards}, the complexity scales with array size, making the approach impractical for large-scale MIMO systems.

\begin{figure}
    \centering
    \includegraphics[width=1\linewidth]{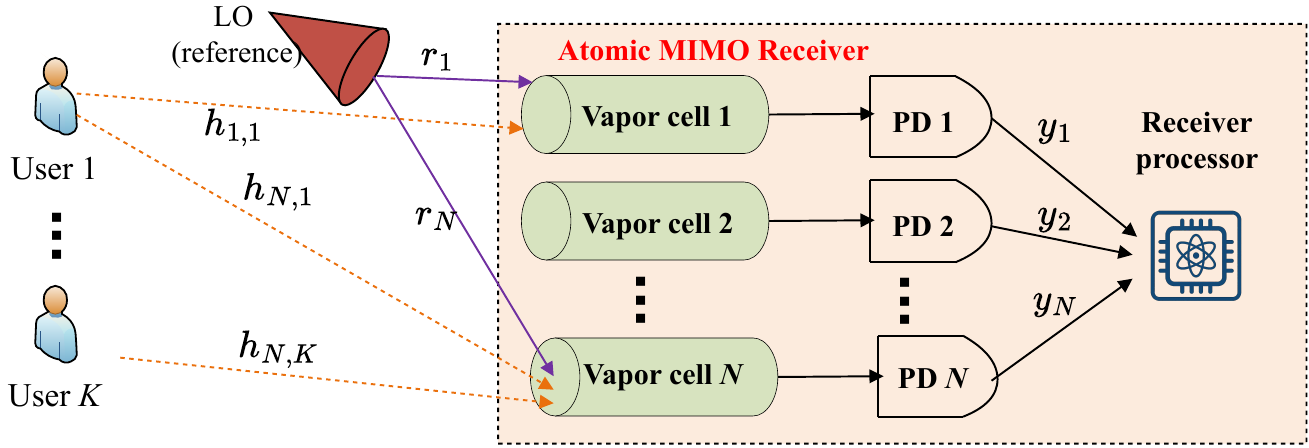}
    \caption{System diagram of the atomic MU-MIMO receiver.}
    \vspace{-5mm}
    \label{fig:atomic}
\end{figure}

Modern wireless network architectures, such as open-radio access network~\cite{bonati2021intelligence}, support artificial intelligence (AI)-driven processing. State-of-the-art sequence AI models have been shown to have in-context learning (ICL) capabilities. Such models receive a prompt containing a few input-output examples, and infer the output for a new input based solely on this context, without updating its internal weights~\cite{garg2022can,zecchin2023context}. ICL has been recently applied to detection tasks for nonlinear MIMO links, where conventional methods struggle due to model mismatch or computational complexity~\cite{zecchin2024cell,song2024transformer,song2024neuromorphic,song2025context,song2025turbo}.

In this work, we introduce a CSI-free symbol detection framework for atomic MIMO receivers based on ICL. The approach uses a small set of pilot-response pairs to directly infer transmitted symbols from nonlinear magnitude-only measurements produced by an atomic receiver. Unlike conventional two-step detection pipelines, the proposed ICL framework offers the following advantages: \emph{(i)} it eliminates the need for explicit channel estimation, thereby avoiding errors associated with inaccurate CSI; \emph{(ii)} it enables direct one-step inference from measurements to symbols, reducing computational overhead; \emph{(iii)} it supports vectorized processing for scalability across large antenna arrays and multiple users. Simulation results confirm that the ICL-based detector achieves higher accuracy and significantly lower runtime compared to optimization-based baselines.

The rest of the paper is organized as follows: Sec.~\ref{se_model} introduces the system model; Sec.~\ref{se:review} reviews conventional detection methods and presents the proposed ICL-based symbol detection approach; Sec.~\ref{sec:model} describes the implementation and training strategy of the ICL-capable network; Sec.~\ref{se:eval} presents numerical results; and Sec.~\ref{se:conclusion} concludes the paper.

\vspace{-1mm}
\section{System Model}\label{se_model}
\vspace{-1mm}
\subsection{Atomic MIMO Measurement Model}

As shown in Fig. \ref{fig:atomic}, a base station (BS) is equipped with an array of $N$ atomic antennas. Each antenna consists of a vapor cell filled with Rydberg atoms integrated with a photodetector (PD). A multi-user (MU) uplink scenario is studied, where $K$ single-antenna users simultaneously transmit to the BS.

Upon receiving EM signals, the Rydberg atoms interact with the local electric field via electric dipole coupling, inducing coherent transitions between their {ground (g) and excited (e) states}. In a typical Ladder configuration, these interactions enable electromagnetically induced transparency (EIT), where destructive interference creates a `dark state' and an optical transparency window \cite{cui2025towards}. 
At each $n$-th vapor cell, the rate of these transitions is characterized by the Rabi frequency $\RabiOmega_n \in \mathbb{C}$. Specifically, denoting $\dipole \in \mathbb{C}^3$ as the complex electric dipole moment vector of the states transition and $\Efield_n \in \mathbb{C}^3$ as the complex envelope of the electric field at the $n$-th vapor cell, the instantaneous Rabi frequency is given by
\begin{equation}\label{eq:rabi_omega}
    \RabiOmega_n = \frac{1}{\hbar} \dipole^\mathsf{H} \Efield_n,
\end{equation}
where $\hbar$ denotes the reduced Planck constant.

The PD output at each antenna measures the population dynamics resulting from the Rabi oscillations. By analyzing the oscillatory behavior of the optical readout as a function of interaction time, we can determine the magnitude of the Rabi frequency $|\RabiOmega_n|$. This measured magnitude, denoted as $y_n$, serves as the output of the $n$-th atomic antenna \cite{cui2025towards}
\begin{equation}\label{eq:antenna_output_y}
    y_n = |\RabiOmega_n| = \frac{1}{\hbar} |\dipole^\mathsf{H} \Efield_n|.
\end{equation}

Let $s_k \in \mathcal{S}$ represent the complex symbol transmitted by user $k$, drawn from a normalized QAM constellation $\mathcal{S}$ satisfying the power constraint $\mathbb{E}[|s_k|^2] = 1$. Each symbol $s_k$ is modulated onto a carrier at frequency $\omega$.

The total local electric field at the $n$-th vapor cell is composed of signal, local oscillation (LO), and noise components. The signal component's complex envelope, $\Efield^{(\text{sig})}_n \in \mathbb{C}^3$, results from multipath propagation of all user signals, and can be expressed as \cite{cui2025towards}
\begin{equation}\label{eq:em_sig_complex}
    \Efield^{(\text{sig})}_n = \sum_{k=1}^{K} \sum_{l=1}^{L_k} \bm{\epsilon}_{nkl} \rho_{nkl} s_k e^{j\varphi_{nkl}},
\end{equation}
where \( L_k \) is the number of paths from user \( k \); \( \bm{\epsilon}_{nkl} \in \mathbb{C}^3 \) is the complex polarization vector; \( \rho_{nkl} \in \mathbb{R}_+ \) is the path gain; and \( \varphi_{nkl} \in [0, 2\pi) \) denotes the path-dependent phase shift for user $k$ and path $l$.

In the absence of a reference signal, a direct measurement \eqref{eq:antenna_output_y} of the signal \eqref{eq:em_sig_complex} would lack sensitivity to the symbol phase \( \gamma_k \), rendering symbol detection ambiguous. To resolve this issue, a holographic phase-sensing approach is adopted as in~\cite{anderson2020rydberg,cui2025towards,cui2024mimo,xu2025channel}, wherein a LO signal with envelope
\begin{equation}\label{eq:lo_complex}
    \Efield^{(\text{LO})}_n = \bm{\epsilon}'_n \rho'_n s_b e^{j\varphi'_n},
\end{equation}
is introduced via a nearby source. This LO also operates at the carrier frequency $\omega$. Here, parameters \( \bm{\epsilon}'_n \in \mathbb{C}^3 \), \( \rho'_n \in \mathbb{R}_+ \), and \( \varphi'_n \in [0, 2\pi) \) denote the complex polarization vector, path loss, and phase shift of the LO field, respectively, and $s_b=|s_b|e^{j\gamma'}\in\mathbb{C}$ is the known LO signal with $\gamma'$ being the phase.

The total electric field is assumed to be perturbed by additive noise prior to interaction with the atoms, accounting for both external EM interference and quantum shot noise. The total complex envelope of the electric field at the $n$-th vapor cell is thus written as the sum
\begin{equation}\label{eq_sig_components_complex}
    \Efield_n = \Efield^{(\text{sig})}_n + \Efield^{(\text{LO})}_n + \bm{w}_n,
\end{equation}
where $\bm{w}_n \in \mathbb{C}^3$ is modeled as a complex Gaussian additive noise vector, 

Substituting \eqref{eq_sig_components_complex} into \eqref{eq:antenna_output_y}, the measured signal $y_n$ can be expressed as {the non-coherent signal model}
\begin{equation}\label{eq:final_yn}
    y_n = \left| \sum_{k=1}^{K} h_{n,k} s_k + r_n + w_n \right|,
\end{equation}
where $h_{n,k} = \frac{1}{\hbar} \dipole^\mathsf{H} \left( \sum_{l=1}^{L_k} \bm{\epsilon}_{nkl} \rho_{nkl} e^{j\varphi_{nkl}} \right)$ represents the effective user channel for user $k$ at antenna $n$, and $r_n = \frac{1}{\hbar} \dipole^\mathsf{H} \Efield^{(\text{LO})}_n = \frac{1}{\hbar} \dipole^\mathsf{H} \bm{\epsilon}'_n \rho'_n s_b e^{j\varphi'_n}$ denotes the effective LO contribution. The effective scalar noise $w_n = \frac{1}{\hbar} \dipole^\mathsf{H} \bm{w}_n$ is obtained by projecting the complex noise field onto the atomic dipole. Under standard assumptions, one obtains the distribution $w_n \sim \mathcal{CN}(0, \sigma^2)$ \cite{cui2025towards}.
 
Define the transmit vector \( \mathbf{s} = [s_1, \dots, s_K]^\top \) by all $k$ users, the received vector \( \mathbf{y} = [y_1, \dots, y_N]^\top \) across $N$ atomic antennas, reference vector \( \mathbf{r} = [r_1, \dots, r_N]^\top \), and the noise vector \( \mathbf{w} = [w_1, \dots, w_N]^\top \). Denote also the user channel to cell \( n \) by \( \mathbf{h}_n = [h_{n,1}, \dots, h_{n,K}]\in\mathbb{C}^{1\times K} \), and the full channel matrix by \( \mathbf{H} = [\mathbf{h}_1^\top; \dots; \mathbf{h}_N^\top]^\top \in \mathbb{C}^{N \times K} \). Given the derivation above, using \eqref{eq:final_yn}, the overall relationship between transmit signal $\mathbf{s}$ and received signal $\mathbf{y}$ can be written as
\begin{equation}\label{eq:measurement_model}
    \mathbf{y} = \left| \mathbf{H} \mathbf{s} + \mathbf{r} + \mathbf{w} \right|,
\end{equation}
where the absolute value is applied element-wise.

Atomic receivers typically operate with a strong reference signal $\mathbf{r}$. This is ensured by placing the LO source near the vapor-cell array, so that the LO signal dominates the incident electric field. In this regime, the magnitude measurement model \eqref{eq:measurement_model} can be linearized via a first-order Taylor approximation around the strong LO amplitude.

Specifically, expanding the received signal $ y_n = \left| r_n + \mathbf{h}_n^\mathsf{H} \mathbf{s} + w_n \right|$ in \eqref{eq:final_yn} around \( r_n \) using the approximation \( |r_n| \gg |\mathbf{h}_n^\mathsf{H} \mathbf{s} + w_n| \), we obtain \cite{cui2024mimo}
\begin{equation}\label{eq:linearized-model-scalar}
y_n - |r_n| \approx \Re \left\{ e^{-j \angle r_n} \left( \mathbf{h}_n^\mathsf{H} \mathbf{s} + w_n \right) \right\}.
\end{equation}
Finally, stacking all \( N \) linearized approximations \eqref{eq:linearized-model-scalar} gives the linearized measurement
\begin{equation}
    \tilde{\mathbf{y}}=\mathbf{y} - |\mathbf{r}| = \Re \big( \mathbf{D}\mathbf{H} \mathbf{s} \big) + \bar{\mathbf{w}},
    \label{eq:linearized-model-vector}
\end{equation}
where \( \mathbf{D} = \operatorname{diag}(e^{-j \angle{\mathbf{r}}}) \in \mathbb{C}^{N \times N} \) is the diagonal matrix representing the conjugate reference phase at each antenna, and \( \bar{\mathbf{w}} \sim \mathcal{N}(0, \tfrac{\sigma^2}{2} \mathbf{I}) \) is real Gaussian noise. In \eqref{eq:linearized-model-vector} and henceforth, we use the equality sign to simplify the notation.

\subsection{Pilot-Aided Transmission}
We consider a quasi-static channel model where the effective channel \( \mathbf{H} \) remains constant within a coherence block. To facilitate CSI acquisition, each user \( k \) transmits a sequence of \( P \) pilot symbols \( \{ \phi_{p,k} \in \mathcal{S} \}_{p=1}^P \), drawn independently from a common normalized QAM constellation \( \mathcal{S} \). At each pilot index \( p \), let \( \bm{\phi}_p = [\phi_{p,1}, \dots, \phi_{p,K}]^\top \in \mathcal{S}^{K \times 1} \) denote the transmitted symbol vector across all users. The corresponding measurements vectors at the atomic MIMO receiver  \( \mathbf{z}_p \in \mathbb{R}^{N\times1} \) is modeled using \eqref{eq:measurement_model} as
\begin{equation}\label{eq:model-pilot-vector}
    \mathbf{z}_p = \left| \mathbf{H} \bm{\phi}_p + \mathbf{r} + \mathbf{w}_p \right|,
\end{equation}
while using the linearized measurement \eqref{eq:linearized-model-vector} one obtaining
\begin{equation}
    \tilde{\mathbf{z}}_p =  \mathbf{z}_p-|\mathbf{r}| = \Re\big( \mathbf{D} \mathbf{H} \bm{\phi}_p \big) + \bar{\mathbf{w}}_p,
    \label{eq:linearized-model-pilot-vector}
\end{equation}
where \( \mathbf{w}_p \sim \mathcal{CN}(0, \sigma^2 \mathbf{I}) \) denotes the additive complex Gaussian noise and \( \bar{\mathbf{w}}_p \sim \mathcal{N}(0, \tfrac{\sigma^2}{2} \mathbf{I}) \) is the real-valued noise.

Stacking the pilot symbols and corresponding measurements across \( P \) pilot instances yields the matrices \( \bm{\Phi} = [\bm{\phi}_1, \dots, \bm{\phi}_p] \in \mathcal{S}^{K \times P} \), \( \mathbf{Z} = [\mathbf{z}_1, \dots, \mathbf{z}_p] \in \mathbb{R}^{N \times P} \) and \( \tilde{\mathbf{Z}} = [\tilde{\mathbf{z}}_1, \dots, \tilde{\mathbf{z}}_p] \in \mathbb{R}^{N \times P} \). The matrix-form measurement models are then expressed as
\begin{subequations}
\begin{equation}\label{eq:model-pilot-matrix}
    \mathbf{Z} = \left| \mathbf{H} \bm{\Phi} + \mathbf{r} \mathbf{1}_{1 \times P} + \mathbf{W}^{\text{p}} \right|,\text{ and}
\end{equation}
\begin{equation}
    \tilde{\mathbf{Z}} = \Re\left( \mathbf{D} \mathbf{H} \bm{\Phi} \right) + \bar{\mathbf{W}}^{\text{p}},
    \label{eq:linearized-model-pilot-matrix}
\end{equation}
\end{subequations}
where \( \mathbf{W}^{\text{p}} \sim \mathcal{CN}(0, {\sigma^2} \mathbf{I}) \) and \( \bar{\mathbf{W}}^{\text{p}} \sim \mathcal{N}(0, \tfrac{\sigma^2}{2} \mathbf{I}) \).

\section{Symbol Detection for Atomic MU-MIMO Receivers}\label{se:review}

This section presents symbol detection approaches for atomic MU-MIMO receivers. First, the conventional pilot-aided approach is reviewed~\cite{cui2025towards,xu2025channel}, and then we introduce the proposed CSI-free approach based on ICL.

\subsection{Conventional Symbol Detection}\label{sse:conventional}
Conventional methods rely on explicit estimation of the CSI using pilot measurements, followed by data symbol detection.

\emph{1) Optimization-Based Channel Estimation:} Given the known transmit pilots \( \bm{\Phi} \), the goal is to estimate the effective channel matrix \( \mathbf{H} \) from the received pilot measurements. This can be formulated as a least-squares (LS) problem based on either the nonlinear or linearized measurement models.

From the magnitude-only measurement model \eqref{eq:model-pilot-vector}, the estimation of \( \mathbf{H} \) involves solving the nonconvex problem:
\begin{equation}
\hat{\mathbf{H}} = \arg\min_{\mathbf{H} \in \mathbb{C}^{N \times K}} \left\| \left| \mathbf{H} \bm{\Phi} + \mathbf{r} \mathbf{1}_{1 \times P} \right| - \mathbf{Z} \right\|_F^2.
\label{eq:ls_nonlinear_channel}
\end{equation}
Alternatively, using the linearized model \eqref{eq:linearized-model-pilot-matrix}, \( \mathbf{H} \) can be estimated by minimizing the squared error:
\begin{equation}
\hat{\mathbf{H}} = \arg\min_{\mathbf{H} \in \mathbb{C}^{N \times K}} \left\| \tilde{\mathbf{Z}} - \Re\left\{ \mathbf{D} \mathbf{H} \bm{\Phi} \right\} \right\|_F^2.
\label{eq:ls_linear_channel}
\end{equation}

{State-of-the-art algorithms include the biased Gerchberg--Saxton (BGS) algorithm~\cite{cui2025towards}, which targets~\eqref{eq:ls_nonlinear_channel}. While effective, BGS has a computational {complexity of order $O\big(N(PK^2 + K^3 + t_0(KP + K^2))\big)$, with the number of iterations $t_0$ ranging in hundreds [{5}, Fig.~7].} For~\eqref{eq:ls_linear_channel}, projected gradient descent (PGD)~\cite{xu2025channel} is commonly employed. PGD is initialized with a complex Gaussian matrix \( \mathbf{H}^{(0)} \) and iteratively updated as
\begin{equation}
\hat{\mathbf{H}}^{(t)} \leftarrow \hat{\mathbf{H}}^{(t-1)} + 2 \eta \cdot \Re\left\{ \left( \tilde{\mathbf{Z}} - \Re\left\{ \mathbf{D} \hat{\mathbf{H}}^{(t-1)} \bm{\Phi} \right\} \right) \bm{\Phi}^\mathsf{H} \right\} \mathbf{D},
\end{equation}
where \( \eta \) denotes the step size and the superscript \( t \) indicates the iteration index. PGD has a complexity of {order \( O(t_0 N K P) \), with hundreds of required iterations, $t_0$, in typical settings [6, Fig.~3]. Furthermore, PGD may suffer from susceptibility to local minima.}

\emph{2) Optimization-Based Symbol Equalization:} With the estimated channel matrix \( \hat{\mathbf{H}} \), the goal of symbol equalization is to recover the transmitted symbol vector \( \mathbf{s}\),  which leads to a discrete nonlinear LS problem. The maximum likelihood (ML) would incur exponential complexity with respect to the number of users \( K \). Relaxing the discrete constraint \( \mathbf{s} \in \mathcal{S}^K \) to the complex space \( \mathbf{s}^* \in \mathbb{C}^K \) yields the continuous non-convex LS problem
\begin{equation}
    \hat{\mathbf{s}}^* = \arg\min_{\mathbf{s}^* \in \mathbb{C}^K} \left\| \left| \hat{\mathbf{H}} \mathbf{s}^* + \mathbf{r} \right| - \mathbf{y} \right\|_2^2.
    \label{eq:equalization_relaxed}
\end{equation}
This relaxed formulation enables the use of gradient-based iterative solvers such as BGS~\cite{cui2025towards}. The resulting estimate \( \hat{\mathbf{s}}^*\) is then quantized to the nearest constellation $\hat{\mathbf{s}}$.

\begin{figure}[t]
    \centering
    \subfigure[]{
    \begin{minipage}{4cm}
        {\begin{tikzpicture}[font=\small,
block/.style={rectangle,very thin, draw=black!50, top color=myyellow!50!,
   bottom color=white, minimum size=10, drop shadow={shadow scale=1,shadow xshift=1pt,shadow yshift=-1pt}, rounded corners=0.6ex, align=center, font=\fontsize{8pt}{9pt}\selectfont},
arrow/.style={->,>=stealth,thick},
textnode/.style={font=\fontsize{8pt}{9pt}\selectfont}
]

\def\tokendistance{0.6cm}
\def\arrowlength{0.3cm}
\def\textdist{0.3cm}
\useasboundingbox (-1.8cm, 1.2cm) rectangle (2.2cm, -0.5cm);

\node[block,minimum height=0.5cm, minimum width = 3.8cm,anchor=south] at (0,0) (an2)  {ICL-Capable Model};

\draw[arrow,color=gray!50!] ($(an2.north)+(-2.5*\tokendistance,0cm)$) -- ($(an2.north)+(-2.5*\tokendistance,\arrowlength)$);
\draw[arrow,color=gray!50!] ($(an2.north)+(-1.5*\tokendistance,0cm)$) -- ($(an2.north)+(-1.5*\tokendistance,\arrowlength)$);
\node[textnode,anchor=center,color=gray!50!] at ($(an2.north)+(-0.5*\tokendistance,0.5*\arrowlength)$) {$\bm{\cdots}$};
\draw[arrow,color=gray!50!] ($(an2.north)+(0.5*\tokendistance,0cm)$) -- ($(an2.north)+(0.5*\tokendistance,\arrowlength)$);
\draw[arrow,color=gray!50!] ($(an2.north)+(1.5*\tokendistance,0cm)$) -- ($(an2.north)+(1.5*\tokendistance,\arrowlength)$);
\draw[arrow] ($(an2.north)+(2.5*\tokendistance,0cm)$) -- ($(an2.north)+(2.5*\tokendistance,\arrowlength)$);

\draw[arrow] ($(an2.south)+(-2.5*\tokendistance,-\arrowlength)$) -- ($(an2.south)+(-2.5*\tokendistance,0cm)$);
\draw[arrow] ($(an2.south)+(-1.5*\tokendistance,-\arrowlength)$) -- ($(an2.south)+(-1.5*\tokendistance,0cm)$);
\node[textnode,anchor=center] at  ($(an2.south)+(-0.5*\tokendistance,-0.5*\arrowlength)$) {$\bm{\cdots}$};
\draw[arrow] ($(an2.south)+(0.5*\tokendistance,-\arrowlength)$) -- ($(an2.south)+(0.5*\tokendistance,0cm)$);
\draw[arrow] ($(an2.south)+(1.5*\tokendistance,-\arrowlength)$) -- ($(an2.south)+(1.5*\tokendistance,0cm)$);
\draw[arrow] ($(an2.south)+(2.5*\tokendistance,-\arrowlength)$) -- ($(an2.south)+(2.5*\tokendistance,0cm)$);

\node[textnode,anchor=center] at ($(an2.north)+(2.5*\tokendistance,\textdist+\arrowlength)$) {$\hat{\mathbf{s}}^*$};

\node[textnode,anchor=center] at ($(an2.south)+(-2.5*\tokendistance,-\arrowlength-\textdist)$) {$\tilde{\mathbf{z}}_1$};
\node[textnode,anchor=center] at ($(an2.south)+(-1.5*\tokendistance,-\arrowlength-\textdist)$) {$\bm{\phi}_1$};
\node[textnode,anchor=center] at ($(an2.south)+(-0.5*\tokendistance,-\arrowlength-\textdist)$) {$\bm{\cdots}$};
\node[textnode,anchor=center] at ($(an2.south)+(0.5*\tokendistance,-\arrowlength-\textdist)$) {$\tilde{\mathbf{z}}_P$};
\node[textnode,anchor=center] at ($(an2.south)+(1.5*\tokendistance,-\arrowlength-\textdist)$) {$\bm{\phi}_P$};
\node[textnode,anchor=center] at ($(an2.south)+(2.5*\tokendistance,-\arrowlength-\textdist)$) {$\tilde{\mathbf{y}}$};

\end{tikzpicture}}
        \label{fig:transformer}
    \end{minipage}   
    }
    \hspace{-7mm}
    \subfigure[]{
    \begin{minipage}{4.8cm}
        \raggedleft
        {\begin{tikzpicture}[font=\small,draw,
block/.style={rectangle,very thin, draw=black!50, top color=myyellow!50!,
   bottom color=white, minimum size=10, drop shadow={shadow scale=1,shadow xshift=1pt,shadow yshift=-1pt}, rounded corners=0.6ex, align=center, font=\fontsize{8pt}{9pt}\selectfont},
arrow/.style={->,>=stealth,thick},
textnode/.style={font=\fontsize{8pt}{9pt}\selectfont,minimum height=0.1cm, minimum width = 0.1cm}
]

\def\tokendistance{0.6cm}
\def\arrowlength{0.3cm}
\def\textdist{0.3cm}

\useasboundingbox (-2.2cm, 1.2cm) rectangle (1.9cm, -0.9cm);

\node[block,minimum height=0.5cm, minimum width = 3.8cm,anchor=south] at (0,0) (an2)  {ICL-Capable Model};

\draw[arrow] ($(an2.north)+(-2.5*\tokendistance,0cm)$) -- ($(an2.north)+(-2.5*\tokendistance,\arrowlength)$);
\draw[arrow,color=gray!50!] ($(an2.north)+(-1.5*\tokendistance,0cm)$) -- ($(an2.north)+(-1.5*\tokendistance,\arrowlength)$);
\node[textnode,anchor=center,color=gray!50!] at ($(an2.north)+(-0.5*\tokendistance,0.5*\arrowlength)$) {$\bm{\cdots}$};
\draw[arrow] ($(an2.north)+(0.5*\tokendistance,0cm)$) -- ($(an2.north)+(0.5*\tokendistance,\arrowlength)$);
\draw[arrow,color=gray!50!] ($(an2.north)+(1.5*\tokendistance,0cm)$) -- ($(an2.north)+(1.5*\tokendistance,\arrowlength)$);
\draw[arrow] ($(an2.north)+(2.5*\tokendistance,0cm)$) -- ($(an2.north)+(2.5*\tokendistance,\arrowlength)$);

\draw[arrow] ($(an2.south)+(-2.5*\tokendistance,-\arrowlength)$) -- ($(an2.south)+(-2.5*\tokendistance,0cm)$);
\draw[arrow] ($(an2.south)+(-1.5*\tokendistance,-\arrowlength)$) -- ($(an2.south)+(-1.5*\tokendistance,0cm)$);
\node[textnode,anchor=center] at  ($(an2.south)+(-0.5*\tokendistance,-0.5*\arrowlength)$) {$\bm{\cdots}$};
\draw[arrow] ($(an2.south)+(0.5*\tokendistance,-\arrowlength)$) -- ($(an2.south)+(0.5*\tokendistance,0cm)$);
\draw[arrow] ($(an2.south)+(1.5*\tokendistance,-\arrowlength)$) -- ($(an2.south)+(1.5*\tokendistance,0cm)$);
\draw[arrow] ($(an2.south)+(2.5*\tokendistance,-\arrowlength)$) -- ($(an2.south)+(2.5*\tokendistance,0cm)$);

\node[textnode,anchor=center] (o1) at ($(an2.north)+(-2.5*\tokendistance,\textdist+\arrowlength)$) {${\mathbf{o}}_{1}$};
\node[textnode,anchor=center] at ($(an2.north)+(-1.5*\tokendistance,\textdist+\arrowlength)$) {$\bm{\cdots}$};
\node[textnode,anchor=center] at ($(an2.north)+(0.5*\tokendistance,\textdist+\arrowlength)$) {${\mathbf{o}}_{2L-3}$};
\node[textnode,anchor=center] (olast) at ($(an2.north)+(2.5*\tokendistance,\textdist+\arrowlength)$) {${\mathbf{o}}_{2L-1}$};

\draw[dashed] ($(o1.north west)$) rectangle ($(olast.south east)-(0.1cm,0cm)$);
\draw[arrow] ($(olast.east)+(-0.1cm,0cm)$) -- ($(olast.east)+(0.1cm,0cm)$);
\node[textnode,anchor=center, inner sep=0pt,outer sep=0pt] at ($(olast.east)+(0.2cm,0cm)$) {$\mathcal{L}$};

\node[textnode,anchor=center] at ($(an2.south)+(-2.5*\tokendistance,-\arrowlength-\textdist)$) {$\tilde{\mathbf{y}}_1^{\text{tr}}$};
\node[textnode,anchor=center] at ($(an2.south)+(-1.5*\tokendistance,-\arrowlength-\textdist)$) {$\mathbf{s}_1^{\text{tr}}$};
\node[textnode,anchor=center] at ($(an2.south)+(-0.5*\tokendistance,-\arrowlength-\textdist)$) {$\bm{\cdots}$};
\node[textnode,anchor=center] at ($(an2.south)+(0.5*\tokendistance,-\arrowlength-\textdist)$) {$\tilde{\mathbf{y}}_{L-1}^{\text{tr}}$};
\node[textnode,anchor=center] at ($(an2.south)+(1.5*\tokendistance,-\arrowlength-\textdist)$) {$\mathbf{s}_{L-1}^{\text{tr}}$};
\node[textnode,anchor=center] at ($(an2.south)+(2.5*\tokendistance,-\arrowlength-\textdist)$) {$\tilde{\mathbf{y}}_{L}^{\text{tr}}$};

\vspace{-10mm}

\end{tikzpicture}}
        \label{fig:infer}
    \end{minipage}  
    }
    \vspace{-2mm}
    \caption{The (a) inference and (b) training process of the ICL-based detector.} 
    \vspace{-5mm}
    \label{fig:train}
\end{figure}
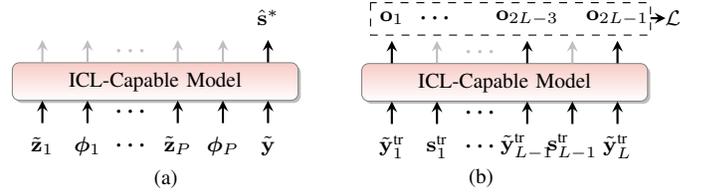
\vspace{-1mm}
\subsection{CSI-Free Symbol Detection via In-Context Learning}

As shown in Fig. \ref{fig:infer}, the proposed ICL approach leverages a sequence model that maps a prompt consisting of contextual information and the received signal to the target output. Specifically, the sequence model receives as input the prompt
\begin{equation}\label{eq:prompt}
    \mathcal{P} = \big\{ \underbrace{\tilde{\mathbf{z}}_1,\, \bm{\phi}_1,\, \dots,\, \tilde{\mathbf{z}}_P,\, \bm{\phi}_P}_{\text{Context } \mathcal{C}},\ \underbrace{\tilde{\mathbf{y}}}_{\text{Target query}} \big\},
\end{equation}
where the \emph{context} \( \mathcal{C}  \) comprises \( P \) pairs of received pilot vectors and their corresponding transmitted symbol vectors. The target query \( \tilde{\mathbf{y}} \) is the received measurement \eqref{eq:linearized-model-vector} associated with the unknown transmitted symbol to be detected.

The objective of ICL is to recover the transmitted symbols associated with the query \( \tilde{\mathbf{y}} \) by leveraging the context examples \( \mathcal{C} \) without explicitly estimating the channel matrix \( \mathbf{H} \). As defined in the next section, this is achieved by learning the inverse mapping of the linearized measurement model in~\eqref{eq:linearized-model-pilot-vector}. The model outputs a soft symbol estimate \( \hat{\mathbf{s}}^* \in \mathbb{C}^{K\times1} \), representing the inferred symbols corresponding to the target measurement.


\vspace{-1mm}
\section{Model Architecture and Training}\label{sec:model}
This section describes the architecture and training strategy of the proposed CSI-free ICL detector.
\vspace{-2mm}
\subsection{Tokenization of Pilot Prompts}\label{subsec:tokenization}
To enable processing by the sequence model, the prompt elements in~\eqref{eq:prompt} are transformed into a sequence of real-valued token vectors \( \{\mathbf{t}_\ell \in \mathbb{R}^{D_T \times 1}\}_{\ell=1}^{2P+1} \), where \( D_T = \max(N, 2K) \). The tokens are defined as
\begin{equation}
\mathbf{t}_\ell =
\begin{cases}
\left[ \tilde{\mathbf{z}}_p^\top,\ \mathbf{0}_{1 \times (D_T - N)} \right]^\top, \quad  \ell = 1,3,\dots,2P-1, \\
\left[ \Re(\bm{\phi}_p^\top),\ \Im(\bm{\phi}_p^\top),\ \mathbf{0}_{1 \times (D_T - 2K)} \right]^\top,\quad  \ell = 2,4,\dots,2P, \\
\left[ \tilde{\mathbf{y}}^\top,\ \mathbf{0}_{1 \times (D_T - N)} \right]^\top, \quad \ell = 2P + 1.
\end{cases}
\end{equation}
Each token \( \mathbf{t}_\ell \) is mapped to a \( D_E \)-dimensional embedding via a learnable projection $\mathbf{W}_{\text{emb}} \in \mathbb{R}^{D_E \times D_T}$ as
$
    \mathbf{e}_\ell^{(0)} = \mathbf{W}_{\text{emb}} \mathbf{t}_\ell, 
$
where \( D_E \) denotes the embedding dimension. Stacking all tokens yields the embedding matrix $ \mathbf{E}^{(0)} = [ \mathbf{e}_1^{(0)}, \dots, \mathbf{e}_{2P+1}^{(0)} ] $. 

\subsection{Transformer-Based Sequence Model}\label{subsec:transformer}
The embedding matrix \( \mathbf{E}^{(0)} \) serves as input to a stack of Transformer decoder blocks. Each block consists of a multi-head self-attention (MHSA) module followed by a position-wise feedforward network (FFN), with residual connections and layer normalization applied at each stage \cite{Vaswani2017AttentionIA}.

Let \( \mathbf{E}^{(n-1)} \in \mathbb{R}^{D_E \times (2P+1)} \) be the input to the \( n \)-th decoder block. Each head \( h \in \{1, \dots, N_H\} \) computes
\begin{equation}
    \mathbf{Q}_h = \mathbf{W}_h^Q \mathbf{E}^{(n-1)}, \quad
    \mathbf{K}_h = \mathbf{W}_h^K \mathbf{E}^{(n-1)}, \quad
    \mathbf{V}_h = \mathbf{W}_h^V \mathbf{E}^{(n-1)},
\end{equation}
where \( \mathbf{W}_h^Q, \mathbf{W}_h^K, \mathbf{W}_h^V \in \mathbb{R}^{D_H \times D_E} \) are learnable projection matrices, and \( D_H = D_E / N_H \) is the dimension of each attention head.
The attention matrices are computed as
\begin{equation}
    \mathbf{A}_h = \text{Softmax}\left( \frac{\mathbf{Q}_h^\top \mathbf{K}_h + \mathbf{M}}{\sqrt{D_H}} \right) \mathbf{V}_h^\top,\; \text{for }h=1,\dots,h,
\end{equation}
where \( \mathbf{M} \in \mathbb{R}^{(2P+1) \times (2P+1)} \) is the causal mask defined by \( M_{i,j} = 0 \) for \( j \leq i \), and \( M_{i,j} = -\infty \) otherwise. This mask ensures that each token attends only to previous tokens in the sequence, mimicking the few-shot inference setup of ICL where the query must not influence the context.

The outputs of all attention heads are concatenated and projected to form the MHSA output:
\begin{equation}
    \mathbf{A} = \mathbf{W}_{\text{MHSA}} \left[ \mathbf{A}_1; \dots; \mathbf{A}_{N_H} \right],
\end{equation}
where \( \mathbf{W}_{\text{MHSA}} \in \mathbb{R}^{D_E \times D_E} \) is a learned projection matrix, and \( [\cdot;\dots;\cdot] \) denotes row-wise concatenation. A residual connection and layer normalization are applied, yielding
\begin{equation}
    \mathbf{E}' = \text{LayerNorm}\left( \mathbf{A} + \mathbf{E}^{(n-1)} \right).
\end{equation}
The FFN applies two linear layers with a nonlinearity, i.e.,
\begin{equation}
    \mathbf{E}^{(n)} = \text{LayerNorm}( \mathbf{W}_2 \text{GeLU}( \mathbf{W}_1 \mathbf{E}' + \bm{\epsilon}_1 ) + \bm{\epsilon}_2 + \mathbf{E}' ),
\end{equation}
where \( \mathbf{W}_1 \in \mathbb{R}^{D_F \times D_E} \), \( \mathbf{W}_2 \in \mathbb{R}^{D_E \times D_F} \), and \( \bm{\epsilon}_1, \bm{\epsilon}_2 \) are bias terms.

The embeddings produced by the last Transformer decoder layer i.e., $\mathbf{E}^{(N_L)} = [ \mathbf{e}_1^{(N_L)},\, \mathbf{e}_2^{(N_L)},\, \dots,\, \mathbf{e}_{2P+1}^{(N_L)} ]$, are projected into a complex-valued output tokens \( \mathbf{o}_\ell \in \mathbb{C}^K \) using a complex-valued output layer:
\begin{equation}
    \mathbf{o}_\ell = \mathbf{W}_{\text{out}} \mathbf{e}_{\ell}^{(N_L)}, \quad \ell = 1, \dots, 2P+1,
\end{equation}
where \( \mathbf{W}_{\text{out}} \in \mathbb{C}^{K \times D_E} \) is a trainable complex-valued projection matrix. Finally, the last output token corresponds to the detected symbol vector, i.e., $\hat{\mathbf{s}}^* = \mathbf{o}_{2P+1}$.

{The complexity to process the context and the first target query is of order $O(N_L P^2 D_E)$, while processing each subsequent target query within the same coherence block is of complexity order $O(N_L P D_E)$ using the key-value caching technique as in \cite{song2025turbo}, as the $2P$ context elements' representations are computed only once and re-used for all subsequent queries. This complexity can be significantly lower than that of the referenced schemes described in Sec.~\ref{sse:conventional}, since the non-parallelizable dimension, i.e., the number of layers $N_L$ is typically much smaller than the number of iterations $t_0$ required by BGS and PGD algorithms.}
\vspace{-2mm}
\subsection{Pre-Training Procedure}\label{subsec:pretraining}
To train the ICL-based detector, we first generate training channels drawn i.i.d. from a pre-training channel distribution $\mathcal{T}_{\text{tr}}=\{\tau_n=(\mathbf{H}_n,\sigma_n^2)\}_{n=1}^{N_{\text{tr}}}\sim \mathcal{D}^{\otimes N_{\text{tr}}}$. For each channel parameters set $\tau_n$, we construct a number of length-$(2L)$ training prompts $\mathcal{P}_{\text{tr}} = \left\{  \tilde{\mathbf{y}}_1^{\text{tr}}, \mathbf{s}_1^{\text{tr}},\dots,\tilde{\mathbf{y}}_{L}^{\text{tr}},\mathbf{s}_{L}^{\text{tr}} \right\}$ consisting $L$ pairs of linearized measurement \( \tilde{\mathbf{y}}_i^{\text{tr}} \in \mathbb{R}^N \) and their corresponding transmitted symbol vector \( \mathbf{s}_i^{\text{tr}} \in \mathcal{S}^K \) generated according to \eqref{eq:measurement_model} and \eqref{eq:linearized-model-vector}. The total length \( L \) is chosen to exceed all possible context lengths \( P \) that may be encountered at inference time.

Each prompt is processed using the procedure in Sec.~\ref{subsec:tokenization}. As shown in Fig. \ref{fig:train}, the model then produces a sequence of complex-valued output vectors \( \{ \mathbf{o}_\ell \in \mathbb{C}^K \}_{\ell=1}^{2L} \), where every odd-indexed output \( \mathbf{o}_{2i-1} \) corresponds to the predicted symbol for the \( i \)-th measurement-symbol pair in the prompt.

We optimize the parameters of the ICL-based detector by minimizing the average mean squared error (MSE) loss over 
\begin{equation}
    \mathcal{L}_\theta = \mathbb{E}_{\tau\sim\mathcal{U}(\mathcal{T_{\text{tr}}}), \mathcal{P}_{\text{tr}}\sim \mathcal{D}_{\mathcal{P}|\tau}}\left[\frac{1}{L} \sum_{i=1}^{L} \left\| \mathbf{o}_{2i-1} - \mathbf{s}_i^{\text{tr}}\right\|_2^2\right],
\end{equation}
where $\mathcal{U}(\cdot)$ denotes uniform distribution, and \( \theta \) denotes the full set of trainable parameters, including token embedding matrices, Transformer layers, and output projection weights. Once trained, the ICL-based detector can generalize to new prompts of arbitrary context length without further fine-tuning.
\vspace{-4mm}
\section{Numerical Evaluation}\label{se:eval}

We evaluate the performance of the proposed CSI-free ICL-based symbol detector in an MU-MIMO uplink scenario with \( K \) single-antenna users. The system employs 4-QAM modulation, and the BS is equipped with \( N = 36 \) atomic antennas. The ICL detector is realized using a compact Transformer model with \( N_L = 4 \) layers and embedding dimension \( D_E = 256 \), comprising approximately 16 million parameters. The model is optimized using the AdamW optimizer with cosine-annealed learning rate scheduling.

We compare the ICL-based detector with two optimization-based approaches. \emph{(i)} \textit{BGS+BGS} employs the BGS algorithm for both channel estimation and nonlinear equalization under the magnitude-only model~\cite{cui2025towards}. \emph{(ii)} \textit{PGD+BGS} uses PGD for channel estimation \cite{xu2025channel}, followed by BGS-based equalization \cite{cui2025towards}. Additionally, a genie-aided benchmark \textit{Perf. CSI + ML} is included, in which the receiver performs ML detection with perfect channel knowledge $\mathbf{H}$.

Fig.~\ref{fig:snr} shows the bit error rate (BER) versus the signal-to-noise ratio (SNR) results. The ICL detector consistently outperforms both BGS+BGS and PGD+BGS across all SNR values. Specifically, ICL achieves approximately 12–14\% lower BER at 0dB to 39–44\% at 7~dB compared to the baselines. Fig.~\ref{fig:usr} presents the BER as a function of the number of users under a fixed SNR of 5~dB. As the number of users $K$ increases from 2 to 8, all methods exhibit performance degradation due to increased interference. Nonetheless, the ICL detector maintains a clear advantage, achieving 19–26\% lower BER than the baselines at $K=8$ to 48–56\% as the number of users decreases to $K=2$. These results underscore the effectiveness and scalability of ICL in learning the underlying input–output mapping from limited pilot data, without relying on explicit channel estimation.

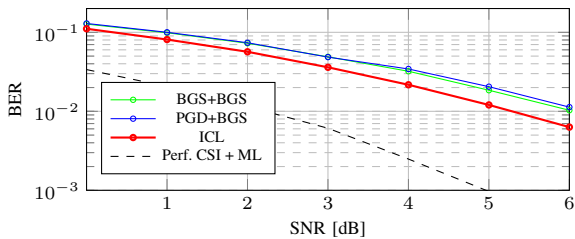
\begin{figure}[t]
    \centering
    \begin{tikzpicture}
\fontsize{7pt}{9pt}\selectfont
    \begin{axis}[
        width=8cm, height=4cm,
        xlabel={SNR [dB]},
        ylabel={BER},
        ymode=log,
        axis lines=box,
        xmin=0, xmax=6,
        ymax=0.2,ymin=1e-3,
        xtick={1,2,3,4,5,6},
        grid=both,
        minor grid style={dashed,gray!50},
        legend pos=south west,
        mark repeat=1,
        legend style={font=\fontsize{6pt}{9pt}\selectfont,row sep=-2pt,yshift=4pt},
    ]
        
        \addplot[
            name path=gs,
            green,
            mark=o,
            mark options={solid},
            mark size=1pt
        ] table[
            col sep=comma,
            x index=0,
            y index=1
        ] {data/4QAM_4x36_pilot_16_gs.csv};
        \addlegendentry{BGS+BGS}
        
        \addplot[
            name path=pgd,
            blue,
            mark=o,
            mark options={solid},
            mark size=1pt
        ] table[
            col sep=comma,
            x index=0,
            y index=1
        ] {data/4QAM_4x36_pilot_16_pgd.csv};
        \addlegendentry{PGD+BGS}

        \addplot[
            name path=icl,
            thick,
            red,
            mark=o,
            mark options={solid},
            mark size=1pt
        ] table[
            col sep=comma,
            x index=0,
            y index=1
        ] {data/4QAM_4x36_pilot_16_icl.csv};
        \addlegendentry{ICL}
        
        \addplot[
            name path=gs_ideal,
            black,
            dashed
        ] table[
            col sep=comma,
            x index=0,
            y index=1
        ] {data/4QAM_4x36_pilot_16_ideal.csv};
        \addlegendentry{Perf. CSI + ML}
    \end{axis}            
\end{tikzpicture}
    \vspace{-3mm}
    \caption{BER versus SNR under 4-QAM modulation with $K=4$ users, $N=36$ receive antennas, and pilot length $P=16$.}
    \vspace{-2mm}\label{fig:snr}
\end{figure}

\begin{figure}[t]
    \centering
    \begin{tikzpicture}
\fontsize{7pt}{9pt}\selectfont
    \begin{axis}[
        width=8cm, height=4cm,
        xlabel={\# Users $K$},
        ylabel={BER},
        ymode=log,
        axis lines=box,
        xmin=2, xmax=8,
        ymax=0.2,ymin=1e-4,
        xtick={2,3,4,5,6,7,8},
        grid=both,
        minor grid style={dashed,gray!50},
        legend pos=south east,
        mark repeat=1,
        legend style={font=\fontsize{6pt}{9pt}\selectfont,row sep=-2pt,yshift=4pt},
    ]
        
        \addplot[
            name path=gs,
            green,
            mark=o,
            mark options={solid},
            mark size=1pt
        ] table[
            col sep=comma,
            x index=0,
            y index=1
        ] {data/4QAM_rx36_SNR_5_pilot_16_gs.csv};
        \addlegendentry{BGS+BGS}
        
        \addplot[
            name path=pgd,
            blue,
            mark=o,
            mark options={solid},
            mark size=1pt
        ] table[
            col sep=comma,
            x index=0,
            y index=1
        ] {data/4QAM_rx36_SNR_5_pilot_16_pgd.csv};
        \addlegendentry{PGD+BGS}

        \addplot[
            name path=icl,
            thick,
            red,
            mark=o,
            mark options={solid},
            mark size=1pt
        ] table[
            col sep=comma,
            x index=0,
            y index=1
        ] {data/4QAM_rx36_SNR_5_pilot_16_icl.csv};
        \addlegendentry{ICL}
        
        \addplot[
            name path=gs_ideal,
            black,
            dashed
        ] table[
            col sep=comma,
            x index=0,
            y index=1
        ] {data/4QAM_rx36_SNR_5_pilot_16_ideal.csv};
        \addlegendentry{Perf. CSI + ML}
    \end{axis}            
\end{tikzpicture}
    \vspace{-3mm}
    \caption{BER versus number of users $K$ at 5 dB SNR under 4-QAM modulation, with $N=36$ receive antennas and pilot length $P=16$.}
    \label{fig:usr}
    \vspace{-5mm}
\end{figure}
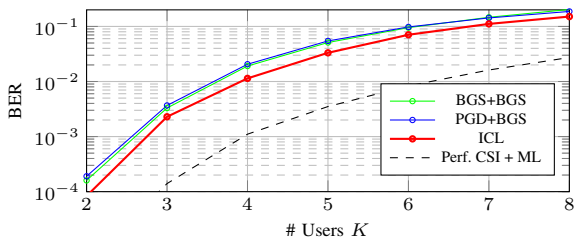

Table~\ref{tab:runtime} shows the average runtime per detection frame under $K=4$, $N=36$ and $p=16$. Both baseline methods are run with the minimum iterations to reach the converged BER in Figs.~\ref{fig:snr} and~\ref{fig:usr}. On CPU, ICL runs 1.6$\times$ faster than BGS+BGS and is comparable to PGD+BGS. Notably, ICL is the only method that is fully parallelizable and natively benefits from standardized GPU acceleration toolkits (e.g., PyTorch-CUDA), whereas GPU implementations of BGS and PGD require manual low-level development \cite{learning2022pycuda}. ICL achieves a 63$\times$ speedup when run on GPU compared to its CPU execution. {Additionally, ICL is 14$\times$ and 3$\times$ faster than the GPU implementation of BGS+BGS and PGD+BGS, respectively. }

\begin{table}[t]
\centering
\caption{Average runtime per detection frame (in milliseconds).}
\label{tab:runtime}
{\scriptsize
\begin{tabular}{lccc}
\toprule
Platform & BGS+BGS & PGD+BGS & ICL \\
\midrule
CPU (Intel Xeon Gold 6542Y) & 7.45 & 4.07 & 4.73 \\
GPU (NVIDIA A100) & 1.043   & 0.246   & \textbf{0.075} \\
\bottomrule
\end{tabular}}
\vspace{-2mm}
\end{table}

\vspace{-2mm}\section{Conclusion}\label{se:conclusion}
We proposed a CSI-free detection framework for atomic MU-MIMO receivers using ICL. ICL eliminates explicit channel estimation by directly mapping pilot-response examples and received symbols to transmitted symbols. Simulation shows that ICL achieves higher accuracy and lower latency compared to state-of-the-art optimization-based approaches across various settings. The findings support the use of AI-native detection for scalable and real-time atomic receivers.

\bibliography{reference}
\end{document}